\documentclass[11pt]{article}

\pdfoutput=1

\usepackage[T1]{fontenc}
\usepackage[latin9]{inputenc}
\usepackage[a4paper]{geometry}
\usepackage[active]{srcltx}
\usepackage{amsmath}
\usepackage{amssymb}
\usepackage{esint}
\usepackage{ulem}

\usepackage{xcolor}

\makeatletter

\usepackage{textcomp}

\pdfoutput=1 

\usepackage{jheppub}

\usepackage{etoolbox}
    
   \patchcmd{\maketitle}{\@fpheader}{}{}{}

\newcommand*\xbar[1]{%
  \hbox{%
    \vbox{%
      \hrule height 0.5pt 
      \kern0.3ex
      \hbox{%
        \kern-0.0em
        \ensuremath{#1}%
        \kern-0.0em
      }%
    }%
  }%
}

\usepackage{amsfonts}

\usepackage{bbm}

\newcommand{\be}{\begin{equation}}
\newcommand{\ee}{\end{equation}}
\newcommand{\bea}{\begin{eqnarray}}
\newcommand{\eea}{\end{eqnarray}}

\title{\boldmath Non-minimal couplings to $U(1)$-gauge fields and asymptotic symmetries}

\author[a,b,c]{Oscar Fuentealba,}
\author[a,d]{Marc Henneaux}
\author[a,e]{and Jules Mas}

\affiliation[a]{Universit\'e Libre de Bruxelles and International Solvay Institutes,\\ ULB-Campus Plaine CP231, B-1050 Brussels, Belgium}
\affiliation[b]{Instituto de Ciencias Exactas y Naturales (ICEN), Universidad Arturo Prat,\\ Playa Brava 3256, 1111346 Iquique, Chile}
\affiliation[c]{Facultad de Ciencias, Universidad Arturo Prat,\\ Avenida Arturo Prat Chac\'on 2120, 1110939 Iquique, Chile}
\affiliation[d]{Coll\`ege de France,  Universit\'e PSL,\\ 11 place Marcelin Berthelot, 75005 Paris, France}
\affiliation[e]{ICFP graduate program, Physics Department, \'Ecole Normale Sup\'erieure, Universit\'e PSL,\\ 24 rue Lhomond, 75005 Paris, France}

\emailAdd{oscar.fuentealba@ulb.be}
\emailAdd{marc.henneaux@ulb.be}
\emailAdd{jules.mas@ens.psl.eu}

\preprint{}

\abstract{We analyse the asymptotic symmetries of electromagnetism non-minimally coupled to scalar fields, with non-minimal couplings of the Fermi type that occur in extended supergravity models.  Our study is carried out at spatial infinity where minimal and non-minimal couplings exhibit very different asymptotic properties: while the former generically cannot be neglected at infinity, the latter can.  Electromagnetic non-minimal couplings are in that respect similar to gravitational minimal couplings, which are also asymptotically subdominant. Because the non-minimally interacting model is asymptotic to the free one, its asymptotic symmetries are the same as the ones of the free theory, i.e., described by angle-dependent $u(1)$ gauge transformations.  We also analyse the duality symmetry and show that it is broken to its compact subgroup by the asymptotic conditions.  Finally, we  consider logarithmic gauge transformations and use them to simplify the symmetry algebra.}

\makeatother

\begin{document}
\maketitle \flushbottom

\section{Introduction}

The asymptotic symmetries of electromagnetism in $4$-dimensional Minkowski space form an infinite-dimensional group parametrized by angle-dependent $u(1)$ transformations \cite{Balachandran:2013wsa,He:2014cra,Kapec:2014zla,Kapec:2015ena,Strominger:2017zoo}, which is somewhat analogous to the BMS group \cite{Bondi:1962px,Sachs:1962wk,Sachs:1962zza}.  While originally found at null infinity, the same symmetry group was exhibited later at spatial infinity \cite{Campiglia:2017mua,Henneaux:2018gfi,Henneaux:2018hdj}.  In particular, the null infinity matching conditions were proved in \cite{Henneaux:2018gfi,Henneaux:2018hdj} to be equivalent to appropriate parity conditions on the leading orders of the Cauchy data in an expansion near spatial infinity.

The introduction of minimal couplings to charged massless fields leads to a more complicated state of affairs. The specific examples of the massless complex Klein-Gordon field or of a collection of spin-$1$ gauge fields interacting through the Yang-Mills mechanism were explicitly investigated with different results at null infinity and spatial infinity.   While the infinite-dimensional symmetry survives at null infinity \cite{Strominger:2013lka,Barnich:2013sxa,He:2014cra,Kapec:2014zla,Kapec:2015ena,Strominger:2017zoo,Lu:2019jus}, there is a tension between this infinite-dimensional symmetry and Lorentz invariance at spatial infinity \cite{Tanzi:2020fmt,Tanzi:2021xva,Tanzi:2021prq}. Lorentz invariance actually restricts  the asymptotic conditions in such a way that no internal asymptotic symmetry survives and only the Poincar\'e symmetry is present.

The difficulties originate from the fact that interactions of the minimal coupling type cannot be neglected at infinity, where the theory does not linearize, contrary to what happens for gravity.  This is very easy to see by mere inspection of the covariant derivatives which read, in the example of a complex one-form $v_k$
\be
D_i v_k = \partial_i v_k - ie A_i v_k  - {\Gamma^j}_{ik}v_j
\ee
where  $A_i$ is the electromagnetic vector potential and ${\Gamma^k}_{ij}$ is the Levi-Civita connection.  With the standard boundary conditions that read at large radial distance $r$ 
\be 
A_i = \mathcal O \left(\frac{1}{r} \right), \qquad g_{ij}- \delta_{ij} = \mathcal O \left(\frac{1}{r} \right),
\ee
one finds 
\be
{\Gamma^k}_{ij} = \mathcal O \left(\frac{1}{r^2} \right) .
\ee
Therefore, $\partial_i v_k $ and $ie A_i v_k$ behave in the same way as $r \rightarrow \infty$ while ${\Gamma^k}_{ij}v_k$ is by contrast subleading and disappears to leading order. Electromagnetic covariant derivatives do not reduce to ordinary derivatives at infinity while gravitational covariant derivatives do.  If $\partial_i v_k$ is relevant at infinity, so is $ie A_i v_k$.

The same is true for the Yang-Mills curvatures where ``minimal coupling'' yields $ F^a _{ij} = f^a_{ij} - g {C^a}_{bc} A^b_i A^c_j$. Here, the $f^a_{ij}$'s are the abelian (free) curvatures, $f^a_{ij}= \partial_i A^a_j - \partial_j A^a_i$. The coupling term and the abelian curvature are of same order $\mathcal O\left(\frac{1}{r^2}\right)$, so that the Yang-Mills curvatures do not linearize at infinity and the interactions cannot be neglected.

It turns out that for massless charged fields, or the Yang-Mills field itself, the minimal coupling terms appear non trivially in the surface integral by which the boosts fail to be canonical transformations under the boundary conditions that allow angle-dependent $u(1)$ transformations at infinity.  This explains the situation described above that  the Poincar\'e group and the angle-dependent $u(1)$ group are incompatible \cite{Tanzi:2020fmt,Tanzi:2021xva,Tanzi:2021prq}.  We explicitly review the argument for massless scalar electrodynamics in the light of the asymptotic non-linearization property in Appendix {\bf \ref{AppendixA}}.

It is interesting to note that the same important difference between the asymptotic properties of gravity and Yang-Mills gauge models plays a striking role in at different (but also asymptotic) context, that of the boost problem of Christodoulou and O'Murchadha \cite{Christodoulou:1981}.  As stressed by these authors, while the Einstein equations linearize at infinity, the Yang-Mills equations exhibit a radically different behaviour which does not guarantee that asymptotically flat initial data have a regular development which includes complete spacelike hypersurfaces boosted relative to the initial one.  

There exist other types of couplings to abelian gauge fields that have been considered in the literature. These are of non-minimal Fermi-type and play for instance an important role in extended supergravity models \cite{Cremmer:1977tt,Cremmer:1978ds,Cremmer:1979up,Gaillard:1981rj,Breitenlohner:1987dg,deWit:2001pz}.   The purpose of this paper is to prove that contrary to minimal couplings, these couplings do not spoil the asymptotic infinite-dimensional angle-dependent $u(1)$ symmetry at spatial infinity. Intuitively, this follows from the fact pointed out above that the equations linearize at infinity as the interaction terms are subdominant with respect to the free ones, and we establish that this intuition is indeed correct.

{The supergravity models enjoy the further interesting feature of being invariant under a ``hidden'' duality symmetry  \cite{Cremmer:1977tt,Cremmer:1978ds,Cremmer:1979up}, which we also include in our asymptotic analysis.}

Our paper is organized as follows. We start with the case of a single Maxwell field coupled to scalar fields on the coset manifold $SL(2,\mathbb{R})/SO(2)$, for which the full duality group is  $SL(2, \mathbb{R})$ (Section  {\bf \ref{Sec:OneMF}}).  We show that in the standard one-potential formulation, the difficulties with boost invariance that arise when one adopts asymptotic conditions allowing non-trivial angle-dependent $u(1)$ asymptotic symmetries can be overcome by exactly the same method as in the pure Maxwell case \cite{Henneaux:2018gfi}. This is because the scalar fields do not contribute to the relevant asymptotic analysis. The theory is then both Poincar\'e invariant and invariant under asymptotic angle-dependent $\mathcal O(1)$ gauge transformations, as in the absence of the scalar fields. 

In Section {\bf \ref{Sec:SL2Duality}}, we consider the manifestly duality invariant formalism, which involves two potentials with a corresponding doubling of the asymptotic angle-dependent $u(1)$-symmetries  \cite{Henneaux:2020nxi}.  We show that the scalar fields do not spoil the asymptotic properties.  We also observe that the boundary conditions break the duality group $SL(2,\mathbb{R})$ to its maximal compact subgroup $SO(2)$, which rotates the two angle-dependent $u(1)$'s.  Section {\bf \ref{Sec:Sp2n}} extends then the model to $n$ Maxwell fields non-minimally coupled to scalar fields parametrizing the coset space $Sp(2n, \mathbb{R})/U(n)$, for which the full duality group is $Sp(2n, \mathbb{R})$ \cite{deWit:2001pz}.  This duality group has been argued recently to play a key role in the quantum theory, even in models where it is only a subgroup of it that is explicitly realized (such as maximal supergravity for which it is  $E_7 \subset Sp(56)$) \cite{Kallosh:2024ull}.  Similar results are obtained: $2n$ independent angle-dependent $u(1)$ asymptotic symmetries, one for each electric potential and one for each magnetic potential; and breaking of the duality group $Sp(2n, \mathbb{R})$ to its maximal compact subgroup $U(n)$ by the boundary conditions.  The global $U(n)$ transformations act as expected on the angle-dependent asymptotic symmetries. 

In Section {\bf \ref{Sec:LogGauge}}, we extend the formalism in another direction, by relaxing the boundary conditions to allow gauge transformations that grow logarithmically at spatial infinity, following  \cite{Fuentealba:2023rvf}.  We find again that the non-minimally coupled scalar fields do not modify the asymptotic properties, and that the new logarithmic gauge symmetries can be used to rewrite the symmetry algebra as a direct sum by appropriate nonlinear redefinitions of the generators.

Finally, Section {\bf \ref{Sec:Conclusions}} is devoted to concluding comments.  Appendix {\bf \ref{AppendixA}} completes the analysis by contrasting the asymptotic features of minimal versus non-minimal couplings at spatial infinity, and the corresponding incompatibility versus compatibility of boost invariance with the relaxed boundary conditions allowing angle-dependent asymptotic symmetries.

\section{$SL(2, \mathbb{R})$-case: One-potential formulation }
\label{Sec:OneMF}

We consider first a single Maxwell field coupled to two scalar fields $ \phi$ and $\chi$ parametrizing the coset manifold $SL(2,\mathbb{R})/SO(2)$, for which the full duality group is  $SL(2, \mathbb{R})$. 

\subsection{Action in Hamiltonian form}

The manifestly covariant action describing this system is given by
\begin{equation}
S=S^{s}[\phi,\chi]+S^{v}[\phi,\chi,A_{\mu}]\,,
\end{equation}
where
\begin{align}
S^{s}[\phi,\chi]&=\int d^4x \sqrt{-g}\,g^{\mu\nu}\left(-\frac{1}{2}\partial_{\mu}\phi\partial_{\nu}\phi-\frac{1}{2}e^{2\phi}\partial_{\mu}\chi\partial_{\nu}\chi\right)\,,\\
S^{v}[\phi,\chi,A_{\mu}]&= \int d^4x \left(-\frac{1}{4}\sqrt{-g}e^{-\phi}F_{\mu\nu}F^{\mu\nu}+\frac{1}{8}\chi\epsilon^{\lambda\mu\rho\sigma}F_{\lambda\mu}F_{\rho\sigma}\right)\,.
\end{align}
In the scalar action $S^{s}$, the field $\phi$ denotes the dilaton, while the scalar field $\chi$ stands for the axion. The vector action $S^{v}$ contains non-minimal couplings between the field strength $F_{\mu \nu}$, associated to the gauge potential $A_{\mu}$, and the dilaton and the axion fields.  These couplings are such that the theory is invariant under $SL(2, \mathbb{R})$-duality, which is manifest in the scalar sector, but more subtle (as a Lagrangian symmetry) in the vector sector \cite{Deser:1976iy,Bunster:2010kwr}.

In order to write the action in Hamiltonian form, we compute the conjugate momentum of each dynamical field, that is to say,\footnote{In our convention the Levi-Civita symbol is chosen such that $\epsilon^{123}=-\epsilon^{0123}=1$.}
\begin{equation}
    \pi^i=\sqrt{\gamma}e^{-\phi}F_0^{\ i}-\frac{1}{2}\chi\epsilon^{ijk}F_{jk}\,,
\end{equation}
\begin{equation}
    \pi_{\phi}=\sqrt{\gamma}\dot{\phi}\,, \quad \pi_{\chi}=\sqrt{\gamma}e^{2\phi}\dot{\chi}\,,
\end{equation}
Then, the Hamiltonian actions read
\begin{align}
S^{s}_H[\phi,\chi;\pi_{\phi},\pi_{\chi}]&=\int dtd^3x (\pi_{\phi}\dot{\phi}+\pi_{\chi}\dot{\chi}-\mathcal{H}^{s})+B^s_{\infty}\,,\label{eq:S-scalar-Ham}\\
S^{v}_H[\phi,\chi,A_i;\pi^i, A_0]&=\int dtd^3x (\pi^i\dot{A}_i-\mathcal{H}^{v}-A_0\mathcal G)+B^v_{\infty}\,,
\end{align}
where 
\be
\mathcal G = - \partial_i \pi^i
\ee
and where the scalar and vector Hamiltonians are given by\footnote{The spatial indices are raised and lowered by the three-dimensional spatial metric $\gamma_{ij}$ and its inverse $\gamma^{ij}$, respectively. }
\begin{align}
\mathcal{H}^{s}&=\frac{1}{2\sqrt{\gamma}}\left(\pi_{\phi}^2+e^{-2\phi}\pi_{\chi}^2\right)+\frac{\sqrt{\gamma}}{2}\gamma^{ij}\left(\partial_i\phi\partial_j\phi+e^{2\phi}\partial_i\chi\partial_j\chi\right)\,,\\
\mathcal{H}^{v}&=\frac{1}{2\sqrt{\gamma}}\left(e^{\phi}\pi_i\pi^i+e^{\phi}\chi\epsilon^{ijk}\pi_iF_{jk}\right) +\frac{\sqrt{\gamma}}{4}\left(e^{\phi}\chi^2+e^{-\phi}\right)F_{ij}F^{ij}\,,
\end{align}
respectively. The specific form of the boundary terms $B^s_{\infty}$ and $B^v_{\infty}$ depends on the boundary conditions.  Variation of the action with respect to $A_0$, which appears as a Lagrange multiplier, yields the Gauss constraint $\mathcal G \approx 0$.

\subsection{Boundary conditions}

In Cartesian coordinates, the fall-off of the dilaton and axion fields, together with their corresponding conjugate momenta is given by
\begin{align}
    \phi&=\frac{\xbar \phi }{r}+\mathcal{O}\left(\frac{1}{r^2}\right)\,, \quad\chi=\frac{\xbar \chi}{r}+\mathcal{O}\left(\frac{1}{r^2}\right)\,, \\ \pi_{\phi}&=\frac{\xbar \pi_{\phi}}{r^2}+\mathcal{O}\left(\frac{1}{r^3}\right)\,,\quad\pi_{\chi}=\frac{\xbar \pi_{\chi}}{r^2}+\mathcal{O}\left(\frac{1}{r^3}\right)\,.
\end{align}This $1/r$-fall-off of the fields is characteristic of massless fields.

Finiteness of the kinetic term in the Hamiltonian scalar action requires to impose parity conditions on the leading order fields under the antipodal map\footnote{The antipodal map is denoted by $n^i\rightarrow -n^i$, where $n^i$ is the unitary vector normal to the sphere at spatial infinity. In spherical polar coordinates, this amounts to the map $(\theta,\varphi)\rightarrow (\pi-\theta,\varphi+\pi)$. }. As in \cite{Henneaux:2018mgn}, we choose the leading order of the scalar fields to be even under the antipodal map, namely,
\begin{equation}
    \xbar \phi(-n^i)=\xbar \phi(n^i)\,,\quad \xbar \chi(-n^i)=\xbar \chi(n^i)\,,
\end{equation}
while the leading order fields of the conjugate momenta are chosen to be odd fields on the sphere:
\begin{equation}
    \xbar \pi_{\phi}(-n^i)= -\xbar \pi_{\phi}(n^i)\,,\quad \xbar \pi_{\chi}(-n^i)=-\xbar \pi_{\chi}(n^i)\,.
\end{equation}

While the Lagrangian is invariant under the full $SL(2, \mathbb{R})$ duality group, the boundary conditions imposed on the scalar fields clearly break this  group to its $SO(2)$ subgroup, which is the stability subroup of the zero field configuration $(\phi = 0, \chi=0)$ to which the scalar fields are required to tend asymptotically.

The fall-off of the vector potential and its conjugate momentum read
    \begin{equation}
        A_i=\frac{\xbar {A_i}}{r}+\mathcal{O}\left(\frac{1}{r^2}\right)\,, \quad \pi^i=\frac{\xbar {\pi}^i}{r^2}+\mathcal{O}\left(\frac{1}{r^3}\right). \label{Eq:Decay0}
    \end{equation}
In order to avoid logarithmic divergences in the symplectic structure, we impose the twisted parity conditions introduced in \cite{Henneaux:2018hdj}. In spherical polar coordinates where the fall-off behaviour reads, 
\begin{equation}
 A_r  = \frac{\xbar {A}_r}{r}+\mathcal{O}\left(\frac{1}{r^2}\right)\,, \quad A_A  = \xbar A_A+\mathcal{O}\left(\frac{1}{r}\right), \quad \pi^r= \xbar \pi^r+\mathcal{O}\left(\frac{1}{r}\right) \, , \quad  \pi^A = \frac{\xbar {\pi}^A}{r}+\mathcal{O}\left(\frac{1}{r^2}\right)\,,\label{Eq:Decay10}
\end{equation}
the radial components are requested to obey strict parity conditions
\begin{equation}
    \xbar A_r(-n^i)=-\xbar A_r(n^i)\,,\quad \xbar \pi^r(-n^i)=\xbar \pi^r(n^i)\,, \label{Eq:Decay1a}
\end{equation}
while the angular components of the vector potential are twisted by a total derivative term
\begin{equation}
    \xbar A_A= \xbar A^{\text{even}}_A+\partial_A \lambda\,,\quad \xbar\pi^A(-n^i)=-\xbar \pi^A(n^i)\,,  \label{Eq:Decay1b}
\end{equation}
where $\xbar A^{\text{even}}_A(-n^i)=\xbar A^{\text{even}}_A(n^i)$ and $\lambda(-n^i)=\lambda(n^i)$. 

If one were to impose strict parity conditions on all components of the vector field \cite{Henneaux:1999ct}, one would find that the only non-trival gauge transformations are the global (angle-independent) ones.  Relaxing them by an angle-dependent gauge transformation that is $\mathcal O(1)$ at infinity and generates leading terms in the potential of opposite parity is the first step towards exhibiting the  infinite-dimensional angle-dependent $u(1)$ asymptotic symmetry at spatial infinity \cite{Henneaux:2018gfi}.

The set of boundary conditions is completed by demanding a faster fall-off of Gauss's constraint, i.e.,
\begin{equation}
\partial_i \pi^i=\mathcal{O}(r^{-4})\,,
\end{equation}
which is also required by finiteness of the kinetic term \cite{Henneaux:2018gfi}.

\subsection{Poincar\'e transformations}

The above boundary conditions in which one allows explicitly an angle-dependent $\mathcal O(1)$ gauge transformation term in the asymptotic form of the fields have been tailored to incorporate an infinite-dimensional angle-dependent $u(1)$ symmetry. A central difficulty with these relaxed boundary conditions, however, is that they conflict with Poincar\'e invariance.  This occurs in a subtle way, in the sense that the boundary conditions themselves are Poincar\'e invariant.  The difficulty has to do with an additional condition that a transformation should fulfill to define an ``invariance'', that of leaving the symplectic form strictly invariant. 

This is a subtle question in the present case because the invariance of the Lagrangian density under  Poincar\'e transformations up to a divergence $\partial_\mu k^\mu$ guarantees the invariance of the symplectic form, but only up to a surface term at spatial infinity.  The whole question is then to check that this surface term vanishes with the chosen boundary conditions.  The point is that it does not with (\ref{Eq:Decay0})-(\ref{Eq:Decay1b}).

This can be cured in the pure Maxwell case.  To understand why the above non-minimal couplings do not spoil the curing procedure, we go step-by-step over it, following closely \cite{Henneaux:2018gfi,Henneaux:2018hdj}.

\subsubsection{Poincar\'e transformations of the fields }

The components of the vector fields generating Poincar\'e transformations read, in spherical coordinates, 
\begin{align}
\xi&=br+T\,,\\
\xi^r&=W\,,\\
\xi^A&=Y^A+\frac{\xbar D^A W}{r}\,,
\end{align}
where $\xbar D_A$ stands for the covariant derivative on the sphere with metric $\xbar \gamma_{AB}$. The parameter $b$ generating Lorentz boosts satisfies the equation 
\begin{equation}
\xbar D_A\xbar D_B b+\xbar g_{AB}b=0\,.
\end{equation}
The constant parameter $T$ generates time translations, while the parameter that generates spatial translations $W$ is subject to the same equation as $b$,
\begin{equation}
    \xbar D_A\xbar D_B W+\xbar g_{AB}W=0\,.
\end{equation}
Finally, $Y^A$ is the sphere Killing vector, i.e., it satisfies the Killing equation $\xbar D_AY_B+\xbar D_BY_A=0$.

The Poincar\'e transformations  in phase space can be obtained as follows.  First, one takes the brackets of the canonical variables with $H_{\xi, \xi^k} \equiv \int d^3x \left[\xi (\mathcal H^s + \mathcal H^v) + \xi^k (F_{km} \pi^m + \pi_\phi \partial_k \phi ) \right]$ where the vector field $(\xi, \xi^k)$ is assumed to decrease sufficiently fast at infinity that $H_{\xi, \xi^k}$  is a well-defined generator, with no extra surface term at infinity  \cite{Regge:1974zd}.  Second, one observes that the variations of the canonical variables obtained in this manner are local in space, so that their variations at a point $x$ do not depend on how the vector field $(\xi, \xi^k)$ behaves at infinity.  The same expressions hold then for Poincar\'e transformations, even though in that case $H_{\xi, \xi^k}$ is a functional that still needs to be improved in order to be a well-defined generator.

One finds that the scalar fields and their corresponding conjugate momenta transform as
\begin{align}
\delta_{\xi,\xi^i}\phi&=\frac{\xi}{\sqrt{\gamma}}\pi_{\phi}+\xi^i\partial_i\phi\,,\\
  \delta_{\xi,\xi^i}\chi&=\frac{\xi}{\sqrt{\gamma}} e^{-2\phi}\pi_{\chi}+\xi^i\partial_i\chi\,,\\
\delta_{\xi,\xi^i}\pi_{\phi}&=\frac{\xi}{\sqrt{\gamma}} e^{-2\phi}\pi_{\chi}^2+\partial_i\left(\sqrt{\gamma}\gamma^{ij}\xi\partial_j\phi\right)-\sqrt{\gamma}\gamma^{ij}\xi e^{2\phi}\partial_i\chi\partial_j\chi\nonumber \\
&\quad-\frac{\xi}{2\sqrt{\gamma}}\left( e^{\phi}\pi_i\pi^i - e^{\phi}\chi\epsilon^{ijk}\pi_iF_{jk}\right) -\frac{\sqrt{\gamma}}{4}\xi\left(e^{\phi}\chi^2-e^{-\phi}\right)F_{ij}F^{ij}+\partial_i\left(\xi^i\pi_{\phi}\right)\,,\\
\delta_{\xi,\xi^i}\pi_{\chi}&=\partial_i(\sqrt{\gamma}\gamma^{ij}\xi e^{2\phi}\partial_j\chi)-\frac{\xi}{2\sqrt{\gamma}} e^{\phi}\epsilon^{ijk}\pi_iF_{jk}-\frac{\sqrt{\gamma}}{2}\xi\chi e^{\phi}F_{ij}F^{ij}+\partial_i\left(\xi^i\pi_{\chi}\right)\,.
\end{align}
The transformation laws of the gauge potential and its conjugate momentum are given by
\begin{align}
\delta_{\xi,\xi^i}A_i&=\frac{\xi}{\sqrt{\gamma}} e^{\phi}\pi_i+\frac{\xi}{2\sqrt{\gamma}} e^{\phi}\gamma_{ij}\chi\epsilon^{jkl}F_{kl}+\mathcal{L}_{\xi^i}A_i\,,\\
\delta_{\xi,\xi^i}\pi^i&=-\partial_j\left(\frac{\xi}{\sqrt{\gamma}}e^{\phi}\chi\epsilon^{ijk} \pi_k\right)-\partial_j\left[\sqrt{\gamma}\xi\left(e^{\phi}\chi^2+e^{-\phi}\right)F^{ij}\right]+\mathcal{L}_{\xi^i}\pi^i\,.
\end{align}

These transformations laws can be easily checked to preserve the asymptotic conditions.  Furthemore, they coincide to leading order with the transformations of the uncoupled fields.  The contributions due to the interactions are easily found to be subleading at least by a power of $1/r$.

\subsubsection{Non-invariance of the symplectic form under Lorentz boosts}

Because the contributions to the Poincar\'e transformations coming from the interactions are subleading, the surface term at infinity giving the variation of the symplectic form
\begin{equation}\label{eq:Symplectic-Form-v1}
    \Omega=\int d^3x\left(d_V\pi_{\phi}d_V \phi+d_V\pi_{\chi}d_V\chi+d_V\pi^id_VA_i\right)\,,
\end{equation}
under boosts takes exactly the same form as in the uncoupled theory.  One finds indeed that  the Lie derivative of the symplectic form along the phase space vector $X_\xi$ associated with boosts is given by the surface term
\begin{equation}
\mathcal{L}_{X_\xi}\Omega=d_V\iota_{X_\xi}\Omega=\oint d^2x\sqrt{\xbar {\gamma}}\,d_V\xbar {A}_r\xbar {D}^A(b d_V\xbar {A}_A)\,,
\end{equation}
exactly as in  \cite{Henneaux:2018gfi}, with no additional contributions coming from the  interactions with the scalar sector.  Because of this key property, one can rescue boost invariance in the same way as in the free case, as we will now show.

This is in sharp contrast with minimal couplings, which modify the leading asymptotic terms in the variations of the fields, leading to obstructions,  as explained in Appendix {\bf \ref{AppendixA}}.

\subsubsection{Re-establishing boost invariance}

To re-establish invariance under Lorentz boosts, the method of \cite{Henneaux:2018gfi}  introduces a surface degree of freedom $\xbar \Psi$ at infinity and modifies the symplectic structure by a surface term that involves it,
\begin{equation}
    \Omega_{\text{surface}}=-\oint d^2x \sqrt{\xbar \gamma}\, d_V\xbar A_rd_V \xbar \Psi\,.
\end{equation}

The new field $\xbar \Psi$ can be extended into the bulk, by incorporating it as the leading order coefficient in the fall-off of the time component of the vector potential $A_0$ and adding its corresponding conjugate momentum $\pi^0$, which must be weakly zero. This amounts to consider  the Hamiltonian formulation of the theory as it comes from the strict application of the Dirac procedure, in which one keeps the ``primary constraint''  $\pi^0 \approx 0$ and its Lagrange multiplier which we denote by $\lambda$ (see e.g. \cite{Dirac1967, Henneaux:1992ig}). The  Hamiltonian action principle with $\pi^0$ and $\lambda$ included is  given by
\begin{equation}
S_H=S^{s}_H[\phi,\chi;\pi_{\phi},\pi_{\chi}]+S^{v}_H[\phi,\chi,A_{\mu};\pi^{\mu}; \lambda]\,,
\end{equation}
where $S^{s}_H$ is the same as in \eqref{eq:S-scalar-Ham} and the vector Hamiltonian action $S^{v}_H$ now reads
\begin{align}
S^{v}_H[\phi,\chi,A_{\mu};\pi^{\mu}; \lambda]&=\int dtd^3x \left(\pi^i\dot{A}_i+\pi^0\dot{A}_0-\mathcal{H}^{v}-\lambda\pi^0\right)-\oint dtd^2x\sqrt{\xbar \gamma}\, \xbar A_r\dot{\xbar \Psi}\,,
\end{align}
where
\begin{align}
\mathcal{H}^{v}&=\frac{1}{2\sqrt{\gamma}}\left(e^{\phi}\pi_i\pi^i+e^{\phi}\chi\epsilon^{ijk}\pi_iF_{jk}\right) +\frac{\sqrt{\gamma}}{4}\left(e^{\phi}\chi^2+e^{-\phi}\right)F_{ij}F^{ij}+A_0\mathcal{G}-\partial_i \pi_0 A^i\,.
\end{align}
We have modified for convenience the vector Hamiltonian by adding the constraint term $-\partial_i \pi_0 A^i$, which is of course permissible.
Variation of the action principle with respect to the Lagrange multiplier $\lambda$ enforces the constraints $\pi^0\approx 0$, while Gauss's law appears now as a ``secondary constraint''.

We can further extend the formalism by introducing an independent Lagrange multiplier $\psi$ for the secondary constraint $\mathcal G$, yielding Dirac's extended formulation  that makes all symmetries manifest \cite{Dirac1967, Henneaux:1992ig}.  This amounts to modifying the vector action as
\be
S^{v}_H[\phi,\chi,A_{\mu};\pi^{\mu}; \lambda] \rightarrow S^{v}_E[\phi,\chi,A_{\mu};\pi^{\mu}; \lambda, \psi] = S^{v}_H[\phi,\chi,A_{\mu};\pi^{\mu}; \lambda] - \int dt d^3x \psi \mathcal G \,.
\ee
The two formulations are physically equivalent and one goes from the extended formulation to the non-extended one by imposing the gauge condition $\psi = 0$.

The fall-off (in Cartesian coordinates) of the time component of the vector potential and its conjugate momentum are given by
\begin{equation}
    A_0=\frac{\xbar \Psi}{r}+\mathcal{O}\left(r^{-2}\right)\,,\quad\pi^0=\frac{\xbar \pi_{\Psi}}{r^2}+\mathcal{O}\left(r^{-3}\right)\,, 
\end{equation}
with parity conditions on the leading orders that read
\begin{equation}
    \xbar \Psi(-n^i)=-\xbar \Psi(n^i)\,,\quad \xbar \pi_{\Psi}(-n^i)=\pi_{\Psi}(n^i)\,.
\end{equation}

The Poincar\'e transformation laws are modified in the vector sector of the theory by adding gauge transformations in such a way that the symplectic form is invariant (see next subsection). This is of course also permissible.  Explicitly,  we take
\begin{align}
\delta_{\xi,\xi^i}A_0&=\nabla_i\left(\xi A^i\right)+\xi^i\partial_iA_0\,,\\
\delta_{\xi,\xi^i}A_i&=\frac{\xi}{\sqrt{\gamma}} e^{\phi}\pi_i+\frac{\xi}{2\sqrt{\gamma}} e^{\phi}\gamma_{ij}\chi\epsilon^{jkl}F_{kl}+\partial_i\left(\xi A_0\right)+\mathcal{L}_{\xi^i}A_i\,,\\
\delta_{\xi,\xi^i}\pi^0&=\xi\partial_i \pi^i+\partial_i\left(\xi^i\pi^0\right)\,,\\
\delta_{\xi,\xi^i}\pi^i&=-\partial_j\left(\frac{\xi}{\sqrt{\gamma}}e^{\phi}\chi\epsilon^{ijk} \pi_k\right)-\partial_j\left[\sqrt{\gamma}\xi\left(e^{\phi}\chi^2+e^{-\phi}\right)F^{ij}\right]+\xi\nabla^i \pi^0+\mathcal{L}_{\xi^i}\pi^i\,.
\end{align}

\subsubsection{Poincar\'e charges}

One can check that the new symplectic form
\begin{equation}\label{eq:Symplectic-Form-v2}
    \Omega=\int d^3x\left(d_V\pi_{\phi}d_V \phi+d_V\pi_{\chi}d_V\chi+d_V\pi^id_VA_i+d_V\pi^0d_VA_0\right)-\oint d^2x \sqrt{\xbar \gamma}\, d_V\xbar A_rd_V \xbar A_0\,,
\end{equation}
is then invariant under Lorentz boosts, and thus under the whole Poincar\'e group, i.e.,
\begin{equation}
    \mathcal{L}_{\xi}\Omega=0\,.
\end{equation}
The Poincar\'e canonical generator can then be obtained by direct application of the equation
\begin{equation}
    \iota_{\xi,\xi^i}\Omega=-d_VP_{\xi,\xi^i}\,.
\end{equation}
Then, we obtain that
\begin{equation}
    P_{\xi,\xi^i}=\int d^3x\left(\xi \mathcal{H}+\xi^i \mathcal{H}_i\right)+\oint d^2x \left[b\left(\sqrt{\xbar \gamma}\,\partial_A\xbar A_r \xbar A^A+\xbar \Psi\,\xbar \pi^r\right)+Y^A\left(\xbar A_A\xbar \pi^r+\sqrt{\xbar \gamma}\,\xbar \Psi\partial_A\xbar A_r\right)\right]\,,
\end{equation}
where the energy and momentum densities read
\begin{align}
\mathcal{H}&=\mathcal{H}^{s}+\mathcal{H}^{v}\,,\\
\mathcal{H}_i&=F_{ij}\pi^j-\partial_j\pi^jA_i+\pi^0\partial_i A_0+\pi_{\phi}\partial_i\phi+\pi_{\chi}\partial_i\chi\,,
\end{align}
respectively, with
\begin{align}
\mathcal{H}^{s}&=\frac{1}{2\sqrt{\gamma}}\left(\pi_{\phi}^2+e^{-2\phi}\pi_{\chi}^2\right)+\frac{\sqrt{\gamma}}{2}\gamma^{ij}\left(\partial_i\phi\partial_j\phi+e^{2\phi}\partial_i\chi\partial_j\chi\right)\,,\\
\mathcal{H}^{v}&=\frac{1}{2\sqrt{\gamma}}\left(e^{\phi}\pi_i\pi^i+e^{\phi}\chi\epsilon^{ijk}\pi_iF_{jk}\right) +\frac{\sqrt{\gamma}}{4}\left(e^{\phi}\chi^2+e^{-\phi}\right)F_{ij}F^{ij}+A_0\mathcal{G}-\nabla_i \pi_0 A^i\,.
\end{align}

\subsection{Angle-dependent $u(1)$ asymptotic symmetries}

The asymptotic conditions are also invariant under the gauge transformations generated by the parameters
\begin{align}
    \epsilon=\xbar \epsilon+\mathcal{O}\left(r^{-1}\right)\,,\quad \mu=\frac{\xbar \mu}{r}+\mathcal{O}\left(r^{-2}\right)\,,
\end{align}
with $\xbar \epsilon (-n^i)=\xbar \epsilon (n^i)$ and $\xbar \mu (-n^i)=-\xbar \mu (n^i)$. Transformation laws of the fields read
\begin{equation}
\delta_{\mu}A_0=\mu\,,\quad\delta_{\epsilon}A_i=\partial_i\epsilon\,,
\end{equation}
where the remaining fields do not transform under gauge transformations. The canonical generator of the asymptotic symmetries is then given by
\begin{equation}
    G_{\mu,\epsilon}=\int d^3x \left(\mu \pi^0+\epsilon\mathcal{G}\right)+\oint d^2x\left(\xbar \epsilon\,\xbar \pi^r-\sqrt{\xbar \gamma}\,\xbar \mu \xbar A_r\right)\,.
\end{equation}

The algebra of the asymptotic symmetries is then the semi-direct sum of Poincar\'e and the infinite-dimensional set of $u(1)$ charges, that is to say,
\begin{align}
\big\{ P_{\xi_{1},\xi_{1}^{i}},P_{\xi_{2},\xi_{2}^{i}}\big\} & =P_{\hat{\xi},\hat{\xi}^{i}}\,,\\
\big\{ G_{\mu,\epsilon},P_{\xi,\xi^{i}}\big\} & =G_{\hat{\mu},\hat{\epsilon}}\,,\\
\big\{ G_{\mu_{1},\epsilon_{1}},G_{\mu_{2},\epsilon_{2}}\big\} & =0\,,
\end{align}
where the components of the Poincar\'e vector field transform as
\begin{align}
\hat{\xi} & =\xi_{1}^{i}\partial_{i}\xi_{2}-\xi_{2}^{i}\partial_{i}\xi_{1}\,,\\
\hat{\xi}^{i} & =\xi_{1}^{j}\partial_{j}\xi_{2}^{i}-\xi_{2}^{j}\partial_{j}\xi_{1}^{i}+g^{ij}\left(\xi_{1}\partial_{j}\xi_{2}-\xi_{2}\partial_{j}\xi_{1}\right)\,.
\end{align}
The gauge parameters are boosted and rotated as follows
\begin{equation}
\hat{\xbar\mu} =-Y^{A}\partial_{A}\xbar\mu-\xbar D_{A}\left(b\xbar D^{A}\xbar\epsilon\right)\,,\quad
\hat{\xbar\epsilon}=-Y^{A}\partial_{A}\xbar\epsilon-b\xbar\mu\,.
\end{equation}

\section{$SL(2, \mathbb{R})$-case: Duality-invariant formulation}
\label{Sec:SL2Duality}

\subsection{Action principle and boundary conditions}

We now turn to the description of the duality symmetry, broken to $SO(2)$ by the asymptotic conditions.  In order to exhibit it, we go to the two-potential formulation along the Hamiltonian lines of \cite{Bunster:2010kwr}.

The Gauss's constraint is solved, by introducing a new vector potential $Z_i$ \cite{Deser:1976iy},
\begin{equation}
    \pi^i=-\epsilon^{ijk}\partial_j Z_k\,.
\end{equation}
We assume the absence of sources, i.e., no electric or magnetic charges, which would otherwise need the introduction of electric and magnetic Dirac strings \cite{Deser:1997mz}.  

Then, up to boundary terms, the vector action takes the form
\begin{equation}\label{eq:vector-action-duality-v1}
  S^v_H=\frac{1}{2}\int dt d^3x\left(\epsilon_{ab}B^{ai}\dot{A}^b_i-\frac{1}{\sqrt{\gamma}}G_{ab}(\phi,\chi)B^{ai}B^b_i\right)\,,  
\end{equation}
where the index $a$ takes the values $(1,2)$ and $A^a_i=(A_i,Z_i)$. The field strength associated with the double vector potential $A^a_i$ is given by
\begin{equation}
B^{ai}=\epsilon^{ijk}\partial_jA^a_k\,.
\end{equation}
The matrix
\begin{equation}
 G_{ab}(\phi,\chi)=   \begin{bmatrix}
e^{\phi}\chi^2+e^{-\phi} & -\chi e^{\phi} \\
-\chi e^{\phi} & e^{\phi}
\end{bmatrix}\,,
\end{equation}
is such that the Hamiltonian in \eqref{eq:vector-action-duality-v1} is invariant under $sl(2,\mathbb{R})$ duality transformations of the fields
\begin{align}
\delta_{\varepsilon} \phi&=\varepsilon^{\alpha}\xi^{\phi}_\alpha(\phi,\chi)\,,\qquad \delta_{\varepsilon}\chi=\varepsilon^{\alpha}\xi^{\chi}_\alpha(\phi,\chi)\,,\\
\delta_{\varepsilon}A^a_i&=\varepsilon^{\alpha}(X_{\alpha})^a_{\,\,\, b}A^b_i\,,\quad \delta_{\varepsilon}B^a_i=\varepsilon^{\alpha}(X_{\alpha})^a_{\,\,\, b}B^b_i\,,
\end{align}
where $\alpha,\beta=\pm,0$. The phase space vectors $\xi_\alpha(\phi,\chi)=\xi^{\phi}_\alpha(\phi,\chi)\frac{\partial}{\partial \phi}+\xi^{\chi}_\alpha(\phi,\chi)\frac{\partial}{\partial \chi}$ and the  matrices $X_{\alpha}$ fulfill both the $sl(2,\mathbb{R})$ algebra,
\begin{equation}
[\xi_{\alpha},\xi_{\beta}]=(\alpha-\beta)\xi_{\alpha+\beta}\,,\qquad [X_{\alpha},X_{\beta}]=(\alpha-\beta)X_{\alpha+\beta}\,.
\end{equation}
 Explicit expressions for these quantities can be found for instance in \cite{Bunster:2010kwr}. 

For the scalar fields and their conjugate momenta we will adopt the asymptotic (and parity) conditions previously written. For the gauge potential, we will take the obvious generalization of (\ref{Eq:Decay10}), i.e., 
\begin{equation}
A^a_r=\frac{\xbar A^a_r}{r}+\mathcal{O}\left(r^{-2}\right)\,,\quad A^a_A=\xbar A^a_A+\mathcal{O}\left(r^{-1}\right)\,.
\end{equation}
This implies the following fall-off for the corresponding field strength
\begin{equation}
B^{ar}=\xbar B^{ar}+\mathcal{O}\left(r^{-1}\right)\,,\quad B^{aA}=\frac{\xbar B^{aA}}{r}+\mathcal{O}\left(r^{-2}\right)\,,
\end{equation}
where
\begin{equation}
    \xbar B^{ar}=\sqrt{\xbar \gamma}e^{AB}\partial_A\xbar A^a_B\,,\quad \xbar B^{aA}=\sqrt{\xbar \gamma}e^{AB}\partial_B\xbar A^a_r\,,
\end{equation}
with $\sqrt{\xbar \gamma}\,e^{AB}\equiv \epsilon^{rAB}$.
We will assume the twisted parity conditions of \cite{Henneaux:2020nxi}
\begin{equation}
    \xbar A^a_r(-n^i)=-\xbar A^a_r(n^i)\,,\quad  \xbar A^a_A= \xbar A^{a\text{even}}_A+\partial_A \lambda^a\,,
\end{equation}
where $\lambda^a(-n^i)=\lambda^a(n^i)$, which guarantee the finiteness of the symplectic structure.  Since only $\partial_A \lambda^a$ appears in the asymptotic conditions, one can assume that these gauge functions have no zero mode. 

\subsection{Poincar\'e invariance}
Rewritten in the two-potential formulation, the Poincar\'e transformations of the fields explicitly read
\begin{align}
\delta_{\xi,\xi^i}\phi&=\frac{\xi}{\sqrt{\gamma}}\pi_{\phi}+\xi^i\partial_i\phi\,,\\
  \delta_{\xi,\xi^i}\chi&=\frac{\xi}{\sqrt{\gamma}} e^{-2\phi}\pi_{\chi}+\xi^i\partial_i\chi\,,\\
\delta_{\xi,\xi^i}\pi_{\phi}&=\frac{\xi}{\sqrt{\gamma}} e^{-2\phi}\pi_{\chi}^2+\partial_i\left(\sqrt{\gamma}\gamma^{ij}\xi\partial_j\phi\right)\nonumber\\
&\quad-\sqrt{\gamma}\gamma^{ij}\xi e^{2\phi}\partial_i\chi\partial_j\chi-\frac{\xi}{2\sqrt{\gamma}}\frac{\partial G_{ab}}{\partial \phi}B^{ai}B^b_i+\partial_i\left(\xi^i\pi_{\phi}\right)\,,\\
\delta_{\xi,\xi^i}\pi_{\chi}&=\partial_i(\sqrt{\gamma}\gamma^{ij}\xi e^{2\phi}\partial_j\chi)-\frac{\xi}{2\sqrt{\gamma}}\frac{\partial G_{ab}}{\partial \chi}B^{ai}B^b_i+\partial_i\left(\xi^i\pi_{\chi}\right)\,,\\
\delta_{\xi,\xi^i}A^a_i&=-\frac{\xi}{\sqrt{\gamma}}\epsilon^{ab}G_{bc}B^c_i+\mathcal{L}_{\xi^i}A^a_i\,.
\end{align}
and are easily verified to preserve the asymptotic conditions. 

The variation of the  symplectic form under Poincar\'e transformations takes again the form
\begin{equation}
    \mathcal{L}_{\xi}\Omega=\frac{1}{2}\oint d^2x \sqrt{\xbar \gamma}\,\delta_{ab}d_V \xbar A^a_r\xbar D_A\left(bd_V\xbar A^{bA}\right)\,,
\end{equation}
with no contribution  from the interaction terms, which are subleading.  

The resolution to the Lorentz boost problem -- the fact that  $\mathcal{L}_{\xi}\Omega \not=0$ -- proceeds then as in \cite{Henneaux:2020nxi}.
We consider first the  Hamiltonian action principle with additional degrees of freedom $\xbar \Psi^a$ at infinity,
\begin{equation}
S_H=S^{s}_H[\phi,\chi;\pi_{\phi},\pi_{\chi}]+S^{v}_H[\phi,\chi,A^a_i,\xbar \Psi^a]\,.
\end{equation}
The scalar action $S^{s}_H$ is the same as in \eqref{eq:S-scalar-Ham}, but the vector Hamiltonian action $S^{v}_H$ in \eqref{eq:vector-action-duality-v1} is modified as follows 
\begin{align}
S^{v}_H[\phi,\chi,A^a_i,\Psi^a]&=\frac{1}{2}\int dt d^3x\left(\epsilon_{ab}B^{ai}\dot{A}^b_i-\frac{1}{\sqrt{\gamma}}G_{ab}(\phi,\chi)B^{ai}B^b_i\right)-\frac{1}{2}\oint dtd^2x\sqrt{\xbar \gamma}\,\delta_{ab} \xbar A^a_r\dot{\xbar \Psi}^b\,.
\end{align}

The transformations under Lorentz boosts of the canonical variables are adjusted by new contributions involving $\xbar \Psi^a$, in such a way that the symplectic form corresponding to the extended action,
\begin{equation}
    \Omega=\int  d^3x\left(d_V\pi_{\phi}d_V \phi+d_V\pi_{\chi}d_V\chi+\frac{1}{2}\epsilon_{ab}d_VB^{ai}d_V {A}^b_i\right)-\frac{1}{2}\oint d^2x\sqrt{\xbar \gamma}\,\delta_{ab}d_V \xbar A^a_rd_V {\xbar \Psi}^b\,,
\end{equation}
is invariant under Poincar\'e transformations. 
One finds that the transformation laws of the scalar fields remain unchanged. The ones of the vector potentials, however, are modified by  gauge transformation terms (as always permissible):
\begin{equation}
    \delta_{\xi,\xi^i}A^a_i=-\frac{\xi}{\sqrt{\gamma}}\epsilon^{ab}G_{bc}B^c_i+\partial_i\left(\xi A^a_0\right)+\mathcal{L}_{\xi^i}A^a_i\,.
\end{equation}
As previously, the fields $A^a_0$ can be thought as  extensions into the bulk of the asymptotic boundary degrees of freedom $\xbar \Psi^a$, i.e., 
\begin{equation}
    A^a_0=\frac{1}{r}\xbar \Psi^a+\mathcal{O}\left(r^{-2}\right)\,.
\end{equation}
Its Poincar\'e variation reads
\begin{equation}  \delta_{\xi,\xi^i}A^a_0=\nabla_i\left(\xi A^{ai}\right)+\xi^i\partial_iA^a_0\,,
\end{equation}
from which it follows that
\begin{equation}
    \delta_{b,Y^A}\xbar \Psi^a=\xbar D_A\left(b\xbar A^{aA}\right)+2b\xbar A^a_r+Y^A\partial_A\xbar \Psi^a\,.
\end{equation}

The Poincar\'e canonical generators, which exist because  $\mathcal{L}_{X_\xi}\Omega=d_V(\iota_{X_\xi} \Omega)$ now vanishes, are obtained from $\iota_{X_{\xi}} \Omega = - d_V P_{\xi,\xi^i}$ and found to be 
\begin{equation}
    P_{\xi,\xi^i}=\int d^3x\left(\xi \mathcal{H}+\xi^i\mathcal{H}_i\right)+B_{\xi,\xi^i}\,,\\
\end{equation}
where the boundary term reads
\begin{equation}
    B_{\xi,\xi^i}=\frac{1}{2}\oint d^2x\left[b\left(\epsilon_{ab}\xbar B^{ar}\xbar \Psi^b-\sqrt{\xbar \gamma}\,\delta_{ab}\xbar A^{aA}\partial_A\xbar A^b_r\right)-\sqrt{\xbar \gamma}\,Y^A\left(\epsilon_{ab}e^{BC}\xbar A^a_A\partial_B \xbar A^b_C-\delta_{ab}\xbar \Psi^a\partial_A \xbar A^b_r\right)\right]\,,
\end{equation}
and the energy and momentum densities are given by
\begin{align}
    \mathcal{H}&=\frac{1}{2\sqrt{\gamma}}\left(\pi_{\phi}^2+e^{-2\phi}\pi_{\chi}^2\right)+\frac{\sqrt{\gamma}}{2}\gamma^{ij}\left(\partial_i\phi\partial_j\phi+e^{2\phi}\partial_i\chi\partial_j\chi\right)+\frac{1}{2\sqrt{\gamma}}G_{ab}(\phi,\chi)B^{ai}B^b_i\,,\\
\mathcal{H}_i&=\pi_{\phi}\partial_i\phi+\pi_{\chi}\partial_i\chi+\frac{1}{2}\epsilon_{ab}\epsilon_{ijk}B^{aj}B^{bk}\,.
\end{align}

\subsection{Asymptotic electric and magnetic angle-dependent symmetries}
In the two-potential formulation, the asymptotic conditions are  invariant under two independent sets of angle-dependent $u(1)$ gauge transformations, one ``electric'' and one ``magnetic''. These are described by the parameters
\begin{align}
    \epsilon^a=\xbar \epsilon^a+\mathcal{O}\left(r^{-1}\right)\,,\quad \mu^a=\frac{\xbar \mu^a}{r}+\mathcal{O}\left(r^{-2}\right)\,,
\end{align}
with $\xbar \epsilon^a (-n^i)=\xbar \epsilon^a (n^i)$ and $\xbar \mu^a (-n^i)=-\xbar \mu^a (n^i)$. Transformation laws of the fields read
\begin{equation}
\delta_{\mu}A^a_0=\mu^a\,,\quad\delta_{\epsilon}A^a_i=\partial_i\epsilon^a\,,
\end{equation}
The canonical generators of the asymptotic symmetries are given by
\begin{equation}
    G_{\mu^a,\epsilon^a}=\frac{1}{2}\oint d^2x\left( \epsilon_{ab}\xbar B^{ar}\xbar\epsilon^b -\sqrt{\xbar \gamma}\,\delta_{ab}\xbar \mu^a \xbar A^b_r\right)\,.
\end{equation}
Because the divergence $\partial_i B^{ai}$ of both magnetic fields identically vanishes, the zero mode of the $u(1)$ symmetries is pure gauge in the sourceless context considered here.  This allows one to assume that the gauge parameters $\epsilon^b$ have no zero mode.

\subsection{Duality symmetry}
Finally, we turn to the duality symmetry.

While the Lagrangian is invariant under $SL(2,R)$ duality transformations,  the boundary conditions on the scalar fields are only preserved by the stability subgroup $SO(2)$ of the configuration $\phi = \chi = 0$. These are parametrized by
\begin{equation}
    \varepsilon^0=0\quad \text{and}\quad \varepsilon^+=-\varepsilon^-=\rho\,,
\end{equation}
with $\rho$ constant and read
\begin{align}
\delta_{\rho}\phi&=2\rho\chi\,,\\
\delta_{\rho}\chi&=\rho\left(e^{-2\phi}-\chi^2-1\right)\,,\\
\delta_{\rho}A^a_i&=\rho \epsilon^{ab}A_{bi}\,.
\end{align}

The canonical generator of $SO(2)$-duality rotations is given by
\begin{equation}
    R_{\rho}=\rho\int d^3x\left[2\pi_{\phi}\chi+\pi_{\chi}\left(e^{-2\phi}-\chi^2-1\right) -\frac{1}{2}\delta_{ab}B^{ai}A^b_i\right]-\frac{\rho}{2}\oint d^2x\sqrt{\xbar \gamma}\, \epsilon_{ab}\xbar A^a_r\,\xbar \Psi^b\,.
\end{equation}

\subsection{Symmetry algebra}

The asymptotic symmetry algebra is the semi-direct sum of the Poincar\'e algebra and two sets of infinite-dimensional $u(1)$ algebras, which transform under $SO(2)$-duality rotations. The non-vanishing Poisson brackets of the asymptotic symmetry algebra read explicitly
\begin{align}
\big\{ P_{\xi_{1},\xi_{1}^{i}},P_{\xi_{2},\xi_{2}^{i}}\big\} & =P_{\hat{\xi},\hat{\xi}^{i}}\,,\\
\big\{ G_{\mu,\epsilon},P_{\xi,\xi^{i}}\big\} & =G_{\hat{\mu},\hat{\epsilon}}\,,\\
\big\{G_{\mu,\epsilon},R_{\rho}\big\} & =G_{\hat{\mu},\hat{\epsilon}}\,,
\end{align}
where the transformed gauge parameters $ \hat{\mu},\hat{\epsilon}$ are given by
\begin{align}
\hat{\xbar\mu}^a &=-Y^{A}\partial_{A}\xbar\mu^a-\xbar D_{A}\left(b\xbar D^{A}\xbar\epsilon^a\right)-\rho\epsilon^{ab}\xbar \mu_b\,,\\
\hat{\xbar\epsilon}^a&=-Y^{A}\partial_{A}\xbar\epsilon^a-b\xbar\mu^a-\rho\epsilon^{ab}\xbar \epsilon_b\,,
\end{align}
where the zero mode in $b\xbar\mu^a$ can be projected out. The transformed Poincar\'e generators $\hat{\xi},\hat{\xi}^{i}$ are given by the standard expression.

\section{Generalization to $Sp(2n,\mathbb{R})$}
\label{Sec:Sp2n}

Our results can be generalized to the system of $2n$ gauge fields $A^M_i$ and scalar fields $\varphi^{\Gamma}$ with the appropiate non-minimal couplings that makes it invariant under $Sp(2n,R)$. We give directly the final results without repeating the explicit derivations.  We follow the notations of \cite{Bunster:2010kwr}.

The respective actions of the scalar and vector sectors read
\begin{align}
S^{s}_H[\varphi^{\Gamma};\pi_{\Gamma}]&=\int dtd^3x \left(\pi_{\Gamma}\dot{\varphi}^{\Gamma}-\frac{1}{2\sqrt{\gamma}}M^{\Gamma\Delta}\pi_{\Gamma}\pi_{\Delta}-\frac{\sqrt{\gamma}}{2}M_{\Gamma\Delta}\gamma^{ij}\partial_i\varphi^{\Gamma}\partial_j\varphi^{\Delta}\right)\,,\label{eq:S-scalar-Ham-Sp}\\
S^{v}_H[\varphi^{\Gamma},A^M_i,\Psi^M]&=\frac{1}{2}\int dt d^3x\left(\sigma_{MN}B^{Mi}\dot{A}^N_i-\frac{1}{\sqrt{\gamma}}G_{MN}(\varphi^{\Gamma})B^{Mi}B^N_i\right)\nonumber\\
&\quad-\frac{1}{2}\oint dtd^2x\sqrt{\xbar \gamma}\,\delta_{MN} \xbar A^M_r\dot{\xbar \Psi}^N\,.
\end{align}
where $\Gamma,\Delta=1,\dots,n^2 + n$ and $M,N=1,\dots,2n$.  Here, 
$$ \sigma = \left(\begin{array}{ccccccc} 0 & 1 & 0 & 0 & \cdots & 0 &0 \\ -1 & 0 & 0 & 0 & \cdots & 0&0\\ 0 & 0 & 0 & 1 & \cdots & 0&0\\ 0 & 0 & -1& 0& \cdots & 0&0 \\ \vdots&\vdots&\vdots& \vdots&\vdots & \vdots & \vdots\\ 0&0&0&0 &\cdots & 0 & 1 \\ 0&0&0&0& \cdots &-1 & 0\end{array} \right), $$
$M_{\Gamma \Delta}(\varphi^{\Gamma})$ is the invariant metric on the scalar manifold (which is the coset space $Sp(2n,\mathbb{R})/U(n)$, of dimension $n^2 + n$), and $G_{MN}(\varphi^{\Gamma})$ is given in \cite{Breitenlohner:1987dg}.

The asymptotic conditions are taken to be as in the $Sp(2)$ case for each respective field. These boundary conditions break the duality symmetry group $Sp(2n,\mathbb{R})$ to its $U(n)$ compact subgroup, which is the stability subgroup of the origin (the scalar fields go to zero at infinity).

The symplectic form defined by the action
\begin{equation}
    \Omega=\int  d^3x\left(d_V\pi_{\Delta}d_V \varphi^{\Delta}+\frac{1}{2}\sigma_{MN}d_VB^{Mi}d_V {A}^N_i\right)-\frac{1}{2}\oint d^2x\sqrt{\xbar \gamma}\,\delta_{MN}d_V \xbar A^M_rd_V {\xbar \Psi}^N\,,
\end{equation}
is invariant under Poincar\'e transformations, i.e.,
\begin{equation}
    d_V (\iota_{\xi}\Omega)=0\,.
\end{equation}
The Poincar\'e generators are given by
\begin{equation}
    P_{\xi,\xi^i}=\int d^3x\left(\xi \mathcal{H}+\xi^i\mathcal{H}_i\right)+B_{\xi,\xi^i}\,,\\
\end{equation}
where the boundary term reads
\begin{align}
    B_{\xi,\xi^i}&=\frac{1}{2}\oint d^2x\left[b\left(\sigma_{MN}\xbar B^{Mr}\xbar \Psi^N+\sqrt{\xbar \gamma}\,\delta_{MN}\xbar A^{MA}\partial_A\xbar A^N_r\right)\right.\nonumber\\
    &\qquad \qquad \qquad \left.+Y^A\left(\sigma_{MN}\xbar B^{Mr}\xbar A^{N}_A+\sqrt{\xbar \gamma}\delta_{MN}\xbar \Psi^M\partial_A \xbar A^N_r\right)\right]\,,
\end{align}
and the energy and momentum densities are given by
\begin{align}
    \mathcal{H}&=\frac{1}{2\sqrt{\gamma}}M^{\Gamma\Delta}\pi_{\Gamma}\pi_{\Delta}+\frac{\sqrt{\gamma}}{2}M_{\Gamma\Delta}\gamma^{ij}\partial_i\varphi^{\Gamma}\partial_j\varphi^{\Delta}+\frac{1}{2\sqrt{\gamma}}G_{MN}(\varphi^{\Gamma})B^{Mi}B^N_i\,,\\
\mathcal{H}_i&=\pi_{\Gamma}\partial_i\varphi^{\Gamma}+\frac{1}{2}\sigma_{MN}\epsilon_{ijk}B^{Mj}B^{Nk}\,.
\end{align}

The angle-dependent asymptotic symmetries form a $U(1)^{2n}$ algebra. Their canonical generators read
\begin{equation}
    G_{\mu^M,\epsilon^N}=\frac{1}{2}\oint d^2x\left( \sigma_{MN}\xbar\epsilon^M \xbar B^{Nr}-\sqrt{\xbar \gamma}\,\delta_{MN}\xbar \mu^M \xbar A^N_r\right)\,.
\end{equation}

Similarly, the canonical generators of $U(n)$-duality rotations are given by
\begin{equation}
    R_{\lambda}=\int d^3x\left[\pi_{\Gamma}\delta_{\lambda}\varphi^{\Gamma} -\frac{1}{2}\sigma_{MP}\lambda^{P}_{\,\,\,N}B^{Mi}A^N_i\right]-\frac{1}{2}\oint d^2x\sqrt{\xbar \gamma}\, \delta_{MP}\lambda^{P}_{\,\,\,N}\xbar A^M_r\,\xbar \Psi^N\,,  \label{Eq:DualityGen}
\end{equation}
where the $U(n)$-duality transformation parameter $\lambda^M_{\,\,\,N}$ satisfies
\begin{equation}
    \sigma_{MN}\lambda^{N}_{\,\,\,P}-\sigma_{MP}\lambda^{N}_{\,\,\,N}=0\quad \text{and} \quad \delta_{MN}\lambda^{N}_{\,\,\,P}+\delta_{MP}\lambda^{N}_{\,\,\,N}=0\,. \label{Eq:ForLambda}
\end{equation}
In (\ref{Eq:DualityGen}), $\delta_{\lambda}\varphi^{\Gamma}$ stands for
\be
\delta_{\lambda}\varphi^{\Gamma} = \lambda^M_{\,\,\,N}\zeta_{(z)}^\Gamma(\varphi) \, \quad \lambda^M_{\,\,\,N} = c^{(z)}  \lambda^M_{\,\,\,(z) N} 
\ee
where $\lambda^M_{\,\,\,(z) N}$ form a basis of solutions of (\ref{Eq:ForLambda}) ($z = 1, \cdots, n^2$) and $\zeta_{(z)}^\Gamma(\varphi)$ are the vector fields tangent to the scalar manifold generating the compact subgroup $U(n)$.

The algebra has the same structure as in the case of a single Maxwell field, where the transformed asymptotic gauge parameters take now the form
\begin{align}
\hat{\xbar\epsilon}^M&=-Y^{A}\partial_{A}\xbar\epsilon^M-b\xbar\mu^M-\lambda^M_{\,\,\,N}\xbar \epsilon^N\,,\\
\hat{\xbar\mu}^M &=-Y^{A}\partial_{A}\xbar\mu^M-\xbar D_{A}\left(b\xbar D^{A}\xbar\epsilon^M\right)-\lambda^M_{\,\,\,N}\xbar \mu^N\,,
\end{align}
which generalizes the above formulas.

\section{Log-gauge transformations - Rewriting the algebra as a direct sum}
\label{Sec:LogGauge}

\subsection{Asymptotic conditions, charges and algebra}

\subsubsection{Action principle with log-relaxed asymptotic conditions} 
A noticeable feature of the symmetry algebra is that the Lorentz generators are not invariant under asymptotic $\mathcal O(1)$ gauge symmetries (although they are of course invariant under gauge transformations that vanish at infinity).  This is because the Poisson brackets of the corresponding generators with the Lorentz generators are non-zero, due to the fact that the gauge generators are in a non-trivial representation of the Lorentz algebra.  Technically, the dependence under ``improper'' \cite{Benguria:1976in} gauge transformations of the Lorentz generators follows from the surface term that must be included to make them well-defined when the more flexible boundary conditions allowing $\mathcal O(1)$ gauge symmetries at infinity are imposed. This surface term involves the bare potentials.   

The gauge-dependence of the Lorentz generators leads to an ambiguity in the values of the angular momentum and boost generators.  In the standard single-potential formulation, one can redefine the Lorentz generators in a manner that makes them invariant under all (proper and improper) gauge transformations \cite{Fuentealba:2023rvf}.  This is done by adding to the Lorentz generators nonlinear terms given (roughly) by the product of the generators of the improper gauge symmetries by the asymptotic fields $\xbar \Psi$ or $\xbar \Phi$. These fields can furthermore be shown to be equal (on the constraint surface) to the generators of a new type of gauge transformations, namely, gauge transformations that grow logarithmically in $r$ at infinity.

The  approach can be straightforwardly extended to the duality invariant formulation with non-minimal couplings to scalar fields considered in this paper.   Ultimately, this is again because the non-minimal couplings are negligeable (to leading order) at infinity, so that the asymptotic computations of the free theory can be repeated without change.

We will thus only sketch here the procedure to arrive at an ambiguity-free angular momentum, following \cite{Fuentealba:2023rvf}, and considering only  the case of a single Maxwell field (in the double potential formulation).  This is sufficient to convey the key ideas. 

To allow the possibility to perform gauge transformations that behave asymptotically as $\ln r$ (``log-gauge transformations"), we take as asymptotic conditions for the gauge potentials:
\begin{align}
A^a_r&=\frac{\xbar A^a_r}{r}+\mathcal{O}\left(r^{-2}\right)\,,\\
A^a_A&=\ln r\,\partial_A\Phi^a_{\log}+\xbar A^a_A+\mathcal{O}\left(r^{-1}\right)\,,\\
A^a_0&=\frac{\ln r}{r}\Psi^a_{\log}+\frac{\xbar \Psi^a}{r}+\mathcal{O}\left(r^{-2}\right)\,.
\end{align}
This leads to the following fall-off for the field strengths:
\begin{equation}
B^{ar}=\xbar B^{ar}+\mathcal{O}\left(r^{-1}\right)\,,\quad B^{aA}=\frac{\xbar B^{aA}}{r}+\mathcal{O}\left(r^{-2}\right)\,,
\end{equation}
where
\begin{equation}
    \xbar B^{ar}=\sqrt{\xbar \gamma}e^{AB}\partial_A\xbar A^a_B\,,\quad \xbar B^{aA}=\sqrt{\xbar \gamma}e^{AB}\partial_B\left(\xbar A^a_r-\Phi^a_{\log}\right)\,.
\end{equation}
 
We impose as in \cite{Fuentealba:2023rvf} the parity conditions
\begin{align}
 \xbar A^a_r(-n^i)&=-\xbar A^a_r(n^i)\,,\quad  \xbar A^a_A=  \left(\xbar A^{a}_A \right)^{\text{even}}+\partial_A \xbar \Phi^a\,,
\end{align}
and
\begin{align}
\xbar \Phi^a(-n^i)&=\xbar \Phi^a(n^i)\,,\quad \xbar \Psi^a(-n^i)=- \xbar \Psi^a(n^i)\,,\\
 \Phi^a_{\log}(-n^i)&=-\Phi^a_{\log}(n^i)\,,\quad \Psi^a_{\log}(-n^i)= \Psi^a_{\log}(n^i)\,.
\end{align}
We can assume that the even functions $\xbar \Phi^a$ have no zero mode, and hence also the even functions $\Psi^a_{\log}$ which are their conjugates in the action  (\ref{Eq:PhiPsi}) below.

The action principle is modified by a surface term involving the new fields $\Psi^a_{\log}$,
\begin{equation}
S_H=S^{s}_H[\phi,\chi;\pi_{\phi},\pi_{\chi}]+S^{v}_H[\phi,\chi,A^a_i,\xbar \Psi^a,  \Psi^a_{\log}]\,,
\end{equation}
yielding the vector Hamiltonian action $S^{v}_H$
\begin{align}
S^{v}_H[\phi,\chi,A^a_i,\Psi^a]&=\frac{1}{2}\int dt d^3x\left(\epsilon_{ab}B^{ai}\dot{A}^b_i-\frac{1}{\sqrt{\gamma}}G_{ab}(\phi,\chi)B^{ai}B^b_i\right)\nonumber \\
&\quad -\frac{1}{2}\oint dtd^2x\sqrt{\xbar \gamma}\,\delta_{ab} \left(\xbar A^a_r\dot{\xbar \Psi}^b+\xbar \Phi^a \dot{\Psi}^b_{\log}\right)\,. \label{Eq:PhiPsi}
\end{align}

\subsubsection{Symmetries and charges}

The preservation under Poincar\'e transformations of the symplectic form implies that the asymptotic fields should transform as
\begin{align}
\delta_{b,Y^A} \Phi^a_{\log}&=b\Psi^a_{\log}+Y^A\partial_A\Phi^a_{\log}\,,\\
\delta_{b,Y^A}\Psi^a_{\log}&=\xbar D_A\left(b\xbar D^A \Phi^a_{\log}\right)+Y^A\partial_A \Psi^a_{\log}\,,
\end{align}
and
\begin{align}
\delta_{b,Y^A} \xbar \Phi^a&=b\xbar \Psi^a+Y^A\partial_A\xbar \Phi^a\,,\\
\delta_{b,Y^A}\xbar \Psi^a&=\xbar D_A\left(b\xbar A^{aA}\right)+2b\xbar A^a_r+Y^A\partial_A\xbar \Psi^a\, ,
\end{align}
with zero modes that can be projected out in $\delta_{b,Y^A}\xbar \Phi^a$ and $\delta_{b,Y^A}\Psi^a_{\log}$.
The Poincar\'e canonical generators are then given by
\begin{equation}
    P_{\xi,\xi^i}=\int d^3x\left(\xi \mathcal{H}+\xi^i\mathcal{H}_i\right)+B_{\xi,\xi^i}\,,\\
\end{equation}
where the boundary term reads
\begin{align}
B_{\xi,\xi^i}&=\frac{1}{2}\oint d^2x\left[b\left(\epsilon_{ab}\xbar B^{ar}\xbar \Psi^b+\sqrt{\xbar \gamma}\,\delta_{ab}\Psi^{a}_{\log}\xbar \Psi^b+\sqrt{\xbar \gamma}\,\delta_{ab}\xbar A^{aA}\partial_A\xbar A^b_r\right)\right.\nonumber\\
&\qquad \qquad \qquad \left.+Y^A\left(\epsilon_{ab}\xbar B^{ar}\xbar A^b_A+\sqrt{\xbar \gamma}\,\delta_{ab}\Psi^{a}_{\log}\xbar A^b_A+\sqrt{\xbar \gamma}\,\delta_{ab}\xbar \Psi^a\partial_A \xbar A^b_r\right)\right]\,.
\end{align}

The asymptotic conditions are also invariant under gauge transformations generated by the parameters
\begin{align}
    \epsilon^a=\ln r\,\epsilon^a_{\log}+\xbar \epsilon^a+\mathcal{O}\left(r^{-1}\right)\,,\quad \mu^a=\frac{\ln r}{r}\mu^a_{\log}+\frac{\xbar \mu^a}{r}+\mathcal{O}\left(r^{-2}\right)\,,
\end{align}
with
\begin{align}
\xbar \epsilon^a (-n^i)&=\xbar \epsilon^a (n^i)\,,\quad \xbar \mu^a (-n^i)=-\xbar \mu^a (n^i)\,,\\
\epsilon^a_{\log} (-n^i)&=- \epsilon^a_{\log} (n^i)\,,\qquad \mu^a_{\log} (-n^i)= \mu^a_{\log} (n^i)\,.
\end{align}
The transformation laws of the fields read
\begin{equation}
\delta_{\mu}A^a_0=\mu^a\,,\quad\delta_{\epsilon}A^a_i=\partial_i\epsilon^a\,,
\end{equation}
The canonical generator of the asymptotic symmetries is then given by
\begin{align}
G_{\mu^a,\epsilon^a}&=\frac{1}{2}\oint d^2x\left[ \left(\epsilon_{ab}\xbar B^{ar}+\sqrt{\xbar \gamma}\,\delta_{ab}\Psi^a_{\log}\right)\xbar\epsilon^b -\sqrt{\xbar \gamma}\,\delta_{ab}\xbar \mu^a \xbar A^b_r\right]\,,\\
G_{\mu^a_{\log},\epsilon^a_{\log}}&=\frac{1}{2}\oint d^2x \sqrt{\xbar \gamma}\, \delta_{ab}\left( \epsilon^a_{\log}\xbar \Psi^b -\xbar \Phi^a \mu^b_{\log}\right)\,.
\end{align}
The zero mode gauge transformations are proper gauge symmetries with zero charge. We can therefore assume here also that the even gauge parameters $\epsilon^a$ and $\mu^a_{\log}$ have no zero mode and this will be done in the description of the symmetry algebra.

To conclude, we have four independent groups of angle-dependent $U(1)$ symmetries, two in the original $\mathcal O(1)$-sector as above, and two new ones in the log-sector (all of which with harmonic number $\ell >0$).

Finally, the canonical generator of $SO(2)$-duality rotations is given by
\begin{align}
R_{\rho}&=\rho\int d^3x\left[2\pi_{\phi}\chi+\pi_{\chi}\left(e^{-2\phi}-\chi^2-1\right) -\frac{1}{2}\delta_{ab}B^{ai}A^b_i\right]\nonumber\\
&\quad-\frac{\rho}{2}\oint d^2x\sqrt{\xbar \gamma}\, \epsilon_{ab}\left(\xbar A^a_r\,\xbar \Psi^b+\xbar \Phi^a\Psi^b_{\log}\right)\,.
\end{align}

\subsubsection{Symmetry algebra}

The computation of the Poisson bracket algebra of the generators is direct. 

The brackets of the $u(1)$-conserved charges and Poincar\'e charges read
\begin{align}
\big\{ G_{\mu_{\log},\epsilon_{\log}},P_{\xi,\xi^{i}}\big\} & =G_{\hat{\mu}_{\log},\hat{\epsilon}_{\log}}\,,\\
\big\{ G_{\mu,\epsilon},P_{\xi,\xi^{i}}\big\} & =G_{\hat{\mu},\hat{\epsilon}}\,,
\end{align}
where
\begin{align}
\hat{\xbar\epsilon}^a&=-Y^{A}\partial_{A}\xbar\epsilon^a-b\xbar\mu^a\,,\\
\hat{\xbar\mu}^a &=-Y^{A}\partial_{A}\xbar\mu^a-\xbar D_{A}\left(b\xbar D^{A}\xbar\epsilon^a\right)\,,\\
\hat{\epsilon}^a_{\log}&=-Y^{A}\partial_{A}\epsilon^a_{\log}-b\mu^a_{\log}\,,\\
\hat{\mu}^a_{\log} &=-Y^{A}\partial_{A}\mu^a_{\log}-\xbar D_{A}\left(b\xbar D^{A}\epsilon^a_{\log}\right)\,.
\end{align}

The brackets of the $u(1)$-conserved charges with the $SO(2)$-rotation generator are given by
\begin{align}
\big\{G_{\mu,\epsilon},R_{\rho}\big\} & =G_{\hat{\mu},\hat{\epsilon}}\,,\\
\big\{G_{\mu_{\log},\epsilon_{\log}},R_{\rho}\big\} & =G_{\hat{\mu}_{\log},\hat{\epsilon}_{\log}}\,,
\end{align}
where
\begin{align}
\hat{\xbar\epsilon}^a&=-\rho\epsilon^{ab}\,\xbar \epsilon_b\,,\quad 
\hat{\xbar\mu}^a =-\rho\epsilon^{ab}\,\xbar \mu_b\,,\\
\hat{\epsilon}^a_{\log}&=-\rho\epsilon^{ab}\epsilon^{\log}_{b}\,,\quad
\hat{\mu}^a_{\log} =-\rho\epsilon^{ab}\mu^{\log}_{b}\,.
\end{align}

The brackets between the $u(1)$-conserved charges give a centrally extended abelian algebra:
\begin{equation}
 \big\{G_{\varepsilon_1},G_{\varepsilon_2}\big\} =\mathcal{C}_{(\varepsilon_1,\varepsilon_2)}\,,   
\end{equation}
where
\begin{align}
    \mathcal{C}_{(\epsilon,\mu_{\log})}&=-\mathcal{C}_{(\mu_{\log},\epsilon)}=\frac{1}{2}\oint d^2x \sqrt{\xbar \gamma}\delta_{ab}\mu^a_{\log}\xbar \epsilon^b \,,\\
    \mathcal{C}_{(\mu,\epsilon_{\log})}&=-\mathcal{C}_{(\epsilon_{\log},\mu)}=-\frac{1}{2}\oint d^2x \sqrt{\xbar \gamma}\delta_{ab}\xbar \mu^a \epsilon^b_{\log}\,.
\end{align}

\subsection{Decoupling of the soft charges in the algebra -- General considerations}

Denoting  the generators of the homogeneous Lorentz generators, spacetime translations, $\mathcal O(1)$-gauge symmetries, log-gauge symmetries and $SO(2)$-duality rotations respectively by $M_I$, $T_j$, $U^a_{\alpha}$, $L^{\alpha}_a$ and $R$, the algebra of the symmetries has the following structure, 
\begin{align}
    \left[M_I,M_J\right]&=f^K_{IJ}M_K\,,\\
    \left[M_I,T_j\right]&=R^k_{Ij}T_k\,,\\
    \left[M_I,U^a_{\alpha}\right]&=G^{\beta}_{I\alpha}U_{\beta}\,,\\
    \left[M_I,L^{\alpha}_a\right]&=-G^{\alpha}_{I\beta}L^{\beta}_a\,,\\
     \left[R,U^a_{\alpha}\right]&=\epsilon^a_b U^{b}_{\alpha}\,,\\
      \left[R,L^{\alpha}_a\right]&=-\epsilon^b_a L^{\alpha}_b\,,\\
     \left[L^{\alpha}_a,U^b_{\beta}\right]&=\delta^b_a \delta^{\alpha}_{\beta}\,,
\end{align}
where $R$ commutes with Poincar\'e generators.
Written in this way,  the algebra takes exactly the form analysed in \cite{Fuentealba:2023hzq}, where it was shown that the presence of an invertible central charge among a set of generators $\{q^i, p_j \}$ enables one to decouple them from the rest of the algebra.  More precisely, through (nonlinear) redefinitions, one can rewrite the algebra as a direct sum involving as one of its summands the subalgebra generated by the $q^i$'s and $p_j$'s.  This method was adopted first in the context of gravity to provide a supertranslation-independent definition of the angular momentum \cite{Fuentealba:2022xsz}.

For the model considered here, one must redefine the Lorentz and $SO(2)$-duality rotation generators as
\begin{align}
    \tilde{M}_I&=M_I-G^{\beta}_{I\alpha}U^a_{\beta}L^{\beta}_a\,,\\
    \tilde{R}&=R-\epsilon^a_b U^b_{\alpha}L^{\alpha}_a\,.
\end{align}
One then finds that the soft charges $U^a_{\alpha}$ and $L^{\beta}_b$ have vanishing Poisson brackets with the new generators.  Furthermore, these have unchanged brackets among themselves,
\begin{align}
    \left[\tilde{M}_I,\tilde{M}_J\right]&=f^K_{IJ}\tilde{M}_K\,,\\
    \left[\tilde{M}_I,T_j\right]&=R^k_{Ij}T_k\,,\\
    \left[\tilde{M}_I,U^a_{\alpha}\right]&=\left[\tilde{M}_I,L^{\alpha}_a\right]=0\,,\\
    \left[\tilde{R},U^a_{\alpha}\right]&=\left[\tilde{R},L^{\alpha}_a\right]=0\,,\\
     \left[L^{\alpha}_a,U^b_{\beta}\right]&=\delta^b_a \delta^{\alpha}_{\beta}\,.
\end{align}
In order to carry out this computation, the following identities are extremely useful:
\begin{itemize}
    \item From 
\begin{equation}
\left[M_I,\left[M_J,U^a_{\alpha}\right]\right]+\left[M_J,\left[U^a_{\alpha},M_I\right]\right]+\left[U^a_{\alpha},\left[M_I,M_J\right]\right]=0\,,
\end{equation}
we get that
\begin{equation}
    G^{\beta}_{J\alpha}G^{\gamma}_{I\beta}-G^{\beta}_{I\alpha}G^{\gamma}_{J\beta}=f^K_{IJ}G^{\gamma}_{K\alpha}\,.
\end{equation}
 \item From 
\begin{equation}
\left[M_I,\left[M_J,T_k\right]\right]+\left[M_J,\left[T_k,M_I\right]\right]+\left[T_k,\left[M_I,M_J\right]\right]=0\,,
\end{equation}
we get that
\begin{equation}
    R^{k}_{Jj}R^{i}_{Ik}- R^{k}_{Ij}R^{i}_{Jk}=f^K_{IJ}R^{i}_{Kj}\,.
\end{equation}
\end{itemize}

\subsection{Decoupling of the soft charges in the algebra -- Explicit computations}

We now put in practice the general considerations of the previous subsection to the case at hand.

The redefinitions of the symmetry generators can equivalently be viewed as redefinitions of the corresponding symmetry parameters.  Applying the general formulas, one finds that the desired redefinition for the Lorentz and $SO(2)$ generators is implemented by performing simultaneously additional $u(1)$-gauge transformations generated by the following parameters
\begin{align}
    \xbar \epsilon^a_{(b,Y^A,\rho)}&=-b\xbar \Psi^a-Y^A\partial_A \xbar \Phi^a-\rho\epsilon^{ab}\xbar \Phi_b\,,\\
    \xbar \mu^a_{(b,Y^A,\rho)}&=-\xbar D_A\left(b\xbar A^{aA}\right)-Y^A\partial_A\xbar \Psi^a-\rho\epsilon^{ab}\xbar \Psi_b\,,\\
    \epsilon^b_{\log(b,Y^A,\rho)}&=-b\left(\frac{\epsilon^{ab}\xbar B^{r}_a}{\sqrt{\xbar \gamma}}+\Psi^b_{\log}\right)-Y^A\partial_A \xbar A^b_r-\rho \epsilon^{bc}\xbar A_{cr} \,,\\
    \mu^b_{\log(b,Y^A,\rho)}&=-\xbar D_A\left(b\xbar D^A\xbar A^b_r\right)-Y^A\partial_A\left(\frac{\epsilon^{ab}\xbar B^{r}_a}{\sqrt{\xbar \gamma}}+\Psi^b_{\log}\right)-\rho \epsilon^{bc}\left(\frac{\epsilon_{ac}\xbar B^{ar}}{\sqrt{\xbar \gamma}}+\Psi^{\log}_{c}\right)\,.
\end{align}
The generators take then the following form
\begin{align}
    \tilde{P}_{\xi,\xi^i}&=\int d^3x\left(\xi \mathcal{H}+\xi^i\mathcal{H}_i\right)\,,\\
    \tilde{R}_{\rho}&=\rho\int d^3x\left[2\pi_{\phi}\chi+\pi_{\chi}\left(e^{-2\phi}-\chi^2-1\right) -\frac{1}{2}\delta_{ab}B^{ai}A^b_i\right]+\frac{\rho}{2}\oint d^2x\delta_{ab}\xbar B^{ar}\,\xbar \Phi^b\,.
\end{align}
It is interesting to note that in the enlarged context where logarithmic gauge transformations are included, the functional $\int d^3x\left(\xi \mathcal{H}+\xi^i\mathcal{H}_i\right)$ is well-defined as a generator, without surface terms. 

We can explicitly check that the only non-vanishing brackets of the canonical generators are given by $ \tilde{P}_{\xi,\xi^i}=\int d^3x\left(\xi \mathcal{H}+\xi^i\mathcal{H}_i\right)$ is invariant under all gauge transformations, proper and improper.  This automatically implies that it has vanishing brackets with the $u(1)$ generators.
\begin{align}
\big\{ \tilde{P}_{\xi_{1},\xi_{1}^{i}},\tilde{P}_{\xi_{2},\xi_{2}^{i}}\big\} & =\tilde{P}_{\hat{\xi},\hat{\xi}^{i}}\,,\\
\big\{G_{\epsilon_1},G_{\epsilon_2}\big\} & =\mathcal{C}_{(\epsilon_1,\epsilon_2)}\,.
\end{align}
    The vanishing of the Poisson brackets of the Poincar\'e generators with the generators of improper gauge symmetries is in particular obvious, since $ \tilde{P}_{\xi,\xi^i}$ is manifestly invariant under both proper and improper gauge transformations.

\section{Concluding remarks}
\label{Sec:Conclusions}

In this paper, we have investigated the asymptotic structure at spatial infinity of electromagnetism coupled to scalar fields through non-minimal couplings of Fermi type.  We have shown that the rich structure found for the free Maxwell theory is unaffected by these couplings, contrary to minimal couplings which do have a non trivial impact asymptotically.  In the asymptotic context, non minimal electromagnetic couplings are  similar to minimal gravitational couplings.

Our study also shows the interplay between asymptotic symmetries and duality symmetries, which arise as ``hidden symmetries'' in supergravity \cite{Cremmer:1978ds}.  It would be interesting to push the analysis further and consider (conjectured) infinite-dimensional hidden symmetries such as $E_{10}$ \cite{Julia1980,Damour:2002cu}, which could be broken to its ``compact'' subgroup $K(E_{10})$ by the relevant boundary conditions.

\section*{Acknowledgments}
 O.F. is grateful to the Coll\`ege de France for kind hospitality while this work was completed. This work was partially supported by a Marina Solvay Fellowship (O.F.) and by  FNRS-Belgium (conventions FRFC PDRT.1025.14 and IISN 4.4503.15), as well as by funds from the Solvay Family.

\appendix

\section{Conflict between Lorentz invariance and angle-dependent $u(1)$ transformations in the case of minimal coupling (at spatial infinity)}
\label{AppendixA}

In this appendix, we consider the action describing the minimal coupling of a complex scalar field $\phi$ to electromagnetism 
\begin{equation}
    S= \int d^4x \sqrt{-g}\left(-\frac{1}{4}F_{\mu\nu}F^{\mu\nu}-\frac{1}{2}(D_{\mu}\phi)^{\dagger}(D^{\mu}\phi)\right),
\end{equation}
where $D_{\mu}=\partial_{\mu}-ieA_{\mu}$ is the gauge covariant derivative and $\dagger$ refers to the complex conjugate.

The momenta that are conjugate to the vector potential $A_i$ read
\begin{equation}
    \pi^i=\sqrt{\gamma}F_0^{\ i},
\end{equation}
while the ones conjugate to the scalar field $\phi$ and its complex conjugate $\phi^{\dagger}$ are
\begin{equation}
     \pi_{\phi}=\sqrt{\gamma}(D_0\phi)^{\dagger}, \quad \pi_{\phi^{\dagger}}=\sqrt{\gamma}(D_0\phi)=\pi_{\phi}^{\dagger}.
\end{equation}
From them, one can derive the following Hamiltonian action
\begin{equation}
    S_H[\phi,\phi^{\dagger},A_i;\pi_{\phi}, \pi_{\phi}^{\dagger},\pi^i]=\int dtd^3x(\pi_{\phi}\dot{\phi}+\pi_{\phi}^{\dagger}\dot{\phi}^{\dagger}+\pi^i\dot{A}_i-\mathcal{H}-A_0\mathcal{G})\,,
\end{equation}
where $\mathcal{G}=-\partial_i\pi^i+ie(\phi\pi_{\phi}-\phi^{\dagger}\pi_{\phi}^{\dagger})$ and the Hamiltonian density is given by 
\begin{equation}
   \mathcal{H}=\frac{\pi^i\pi_i}{2\sqrt{\gamma}}+\frac{2}{\sqrt{\gamma}}\pi_{\phi}\pi_{\phi}^{\dagger}+\frac{\sqrt{\gamma}}{4}F_{ij}F^{ij}+\frac{\sqrt{\gamma}}{2}(D_i\phi)^{\dagger}(D^i\phi).
\end{equation}
The transformation laws of the dynamical fields under boosts and time translations parametrized by $\xi$  are defined up to a gauge transformation generated by $\zeta$ and read
\begin{equation}
    \delta_{\xi}A_i=\frac{\xi\pi_i}{\sqrt{\gamma}}+\partial_i\zeta, \quad \delta_{\xi}\phi=\frac{2\xi\pi_{\phi}^{\dagger}}{\sqrt{\gamma}}+ie\zeta \phi, \quad \delta_{\xi}\phi^{\dagger}=\frac{2\xi\pi_{\phi}}{\sqrt{\gamma}}-ie\zeta\phi^{\dagger},
\end{equation}
\begin{equation}
    \delta_{\xi}\pi^i=\partial_j(\xi\sqrt{\gamma}F^{ji})-\frac{\xi\sqrt{\gamma}}{2}ie\phi^{\dagger}\partial^i\phi+\frac{\xi\sqrt{\gamma}}{2}ie\phi\partial^i\phi^{\dagger} -e^2\xi\sqrt{\gamma}\phi^{\dagger}\phi A^i,
\end{equation}
\begin{equation}
    \delta_{\xi}\pi_{\phi}=\frac{1}{2}\partial_i(\xi\sqrt{\gamma}(D^i\phi)^{\dagger})+\frac{ie}{2}\xi\sqrt{\gamma}(D^i\phi)^{\dagger}A_i-ie\zeta\pi_{\phi},
\end{equation}
\begin{equation}
    \delta_{\xi}\pi_{\phi}^{\dagger}=\frac{1}{2}\partial_i(\xi\sqrt{\gamma}(D^i\phi))-\frac{ie}{2}\xi\sqrt{\gamma}(D^i\phi)A_i+ie\zeta\pi_{\phi}^{\dagger}.
\end{equation}
\newline
Then, for the vector field $X$ defined by these infinitesimal transformations, one gets
\begin{align}\label{eq:App}
    d_V\iota_X\Omega=&\oint d^2x\sqrt{\xbar{\gamma}}d_V\xbar A_r\xbar{D}^A(b d_V\xbar{A}_A)\nonumber \\ &+\oint d^2x b\sqrt{\xbar{\gamma}}ie\left(\xbar{A}_rd_V\xbar{\phi}^{\dagger}d_V\xbar{\phi}+\frac{1}{2}\xbar{\phi}^{\dagger}d_V\xbar{A}_rd_V\xbar{\phi}-\frac{1}{2}\xbar{\phi}d_V\xbar{A}_rd_V\xbar{\phi}^{\dagger}\right)
\end{align}
so that the scalar field appears also in the surface term that prevents the boosts to be canonical. Therefore, one can not reproduce the same solution as the one provided in the free theory \cite{Henneaux:2018gfi}, contrary to what is happening for the non minimal coupling we considered  in this paper. Specifically, the second line at the right hand side of \eqref{eq:App} cannot be cancelled by a boundary deformation of the symplectic form by using a general ansatz \cite{Tanzi:2021xva}. Thus, in order to get rid of these new extra terms one must impose strict parity conditions on the leading order fields, which freezes the possibility of having infinite-dimensional gauge symmetries \cite{Tanzi:2021xva} (see also \cite{Tanzi:2021prq}).

\end{document}